\documentstyle [prl,aps]{revtex}
\input epsf
\begin{document}

\draft
\title{Endpoint Structure in Beta Decay from 
Coherent Weak-Interaction of the Neutrino}

\author{
J.I. Collar$^{a,b,*}$
}
\address{ 
$^{a}$CERN, EP Division, CH-1211 Geneve 23, Switzerland\\
$^{b}$Department of Physics and Astronomy, University of South Carolina, 
Columbia, SC 29208\\
}
\wideabs{
\maketitle
\begin{abstract}
\widetext
Recent tritium beta decay experiments yield unphysical 
negative best-fit values for the square of the neutrino mass. 
An unidentified bump-like excess of counts few eV below the 
endpoint in the electron energy spectrum has been tentatively 
recognized as the source of this anomaly. It is shown that the 
repulsive potential acting on the emitted antineutrino and 
originating in its coherent weak-interaction with the daughter 
atom may effectively account for this excess.  
\end{abstract}

\pacs{PACS number(s): 
23.40.Bw(-s), 14.60.Pq\\
}
$^{*}$~E-mail: Juan.Collar@cern.ch}

\narrowtext
The evidence for a non-zero electron neutrino mass   
$m_{\nu}\sim 30$ eV/$c^{2}$, obtained by the ITEP group in the 80's 
\cite{1}
from a measurement of the endpoint region in the  $\beta-$decay 
of tritium, has been amply refuted in later experiments 
\cite{2,3,4,5,6,7}. 
These subsequent attempts have nevertheless consistently 
produced unphysical $m^{2}_{\nu}<0$ values, prompting the Particle 
Data Group to devise a special recipe \cite{8} to translate them 
into sensible upper limits to the positive value of $m^{2}_{\nu}$.
Recently, 
a common origin for this anomaly has been independently 
suggested by some of these groups \cite{5,6,7,9}: a broad spike or 
bump-like excess of counts centred 5 to 30 eV below the 
endpoint energy $E_{0}$ in the electron kinetic energy ($E_{e}$) 
spectrum, is able to explain the effect. Best values 
for the position, intensity and spectral shape of this 
bump are somewhat different from one experiment to another. 
This is expected from the low signal-to-background ratio 
close to $E_{0}$ (the first statistically significant data points 
are typically found 5 to 20 eV below the endpoint) and the 
different energy resolutions of the spectrometers 
($\sim$5 to 15 eV at best). The origin of this structure is yet 
unknown and several hypothesis such as residual 
radioactivity generating a monochromatic electron line \cite{6,7} 
or an increased shake-off probability \cite{5} have been ruled out. 
Stephenson \cite{9} has revised the Los Alamos result \cite{3} 
including the competing process of relic neutrino absorption, 
which is expected to generate a weak monochromatic electron 
line at or above $E_{0}$ \cite{10}. In his interpretation, the actual 
contribution from this process would be an essentially constant 
addition to the region  $E_{0}-E_{F}<E_{e}<E_{0}$, where $E_{F}\sim$
few eV is the energy 
of the relic neutrino Fermi sea. Stephenson nevertheless finds 
that the required present-epoch relic neutrino density necessary 
to produce the observed excess is a factor $10^{14}$ larger than the 
$\sim 110~\nu/cm^{3}$   
predicted by standard Big-Bang cosmology.

The possible link between the position and intensity of the 
bump has not been examined yet. Fig. 1 shows their available 
best-values and error bars as listed in refs. \cite{6} (Troitsk), \cite{7} 
(Lawrence Livermore National Laboratory) and \cite{9} (Los Alamos National 
Laboratory). The position (centroid) is defined by an energy  $\varepsilon$
below $E_{0}$ and the intensity by the fraction of the total $\beta-$decay 
strength under the bump. The Troitsk group gives a best-value  $\varepsilon=$7 eV 
with no associated error bar, but elsewhere in \cite{6} they 
define  $\varepsilon\sim$7 - 15 eV. As for the LANL result, the shaded region 
in fig. 1 spans over fits with the proposed relic-absorption 
spectral shape of \cite{9} that give a goodness-of-fit equivalent 
to their earlier attempt \cite{3} at fitting a sharp spike at $\varepsilon=$ 0 
(yielding an intensity $\sim 10^{-9}$ of the total decays). The Mainz 
group \cite{5} does not offer a best value, but the position and 
magnitude of the deviation is reported as "remarkably similar" 
to the LLNL result \cite{7}. The fraction of the decays for which  
$E_{e}>E_{0}-\varepsilon$, 
that is, $f(\varepsilon)=\int_{E_{0}-\varepsilon}^{E_{0}}P(E_{e})dE_{e} / 
\int_{0}^{E_{0}}P(E_{e})dE_{e}$, where $P(E_{e})dE_{e}$ is 
the electron differential kinetic 
energy spectrum, is also depicted in fig. 1. The theoretical 
$P(E_{e})dE_{e}$ is 
calculated following Morita \cite{11} and includes the Coulomb-screening 
correction to the relativistic Fermi function \cite{12,13} and the finite 
deBroglie wavelength correction. $E_{0}=$18575 eV is adopted 
($f(\varepsilon)$
is not very sensitive to a variation of $\sim$20 eV in this value). 
The solid line represents the case $m_{\nu}=$ 0 while the dotted line 
is for $m_{\nu}=$ 5 eV.

The closeness of these experimental best-values and the 
theoretical curve $f(\varepsilon)$, compatible with a small $m_{\nu}$, 
is remarkable; 
there is no self-evident reason why such a finely-tuned correlation 
between the position and intensity of the bump should exist. 
Its presence in all experiments points at a common cause and 
provides an intuitive hint of its origin: antineutrinos emitted 
accompanying electrons with  $E_{0}-\varepsilon<E_{e}<E_{0}$ 
have a small kinetic energy $<\varepsilon$; 
imposing a requirement that antineutrinos always carry a minimum 
amount of total energy, $E_{\nu}>V_{c}$, where $V_{c}\sim$ 
few eV is some repulsive potential 
acting on them, might in principle effectively translate into "lifting"   
$P(E_{e})dE_{e}$ around $E_{e}\sim E_{0}-V_{c}$  
by an amount equivalent to $f(V_{c})$, i.e., as if electron 
emission into $E_{e}>E_{0}-V_{c}$  was energetically forbidden and these electrons 
piled-up at  $E_{e}=E_{0}-V_{c}$. This description is nevertheless shown below to be 
formally inaccurate. 
\begin{figure}[tbp]
\epsfxsize = \hsize \epsfbox{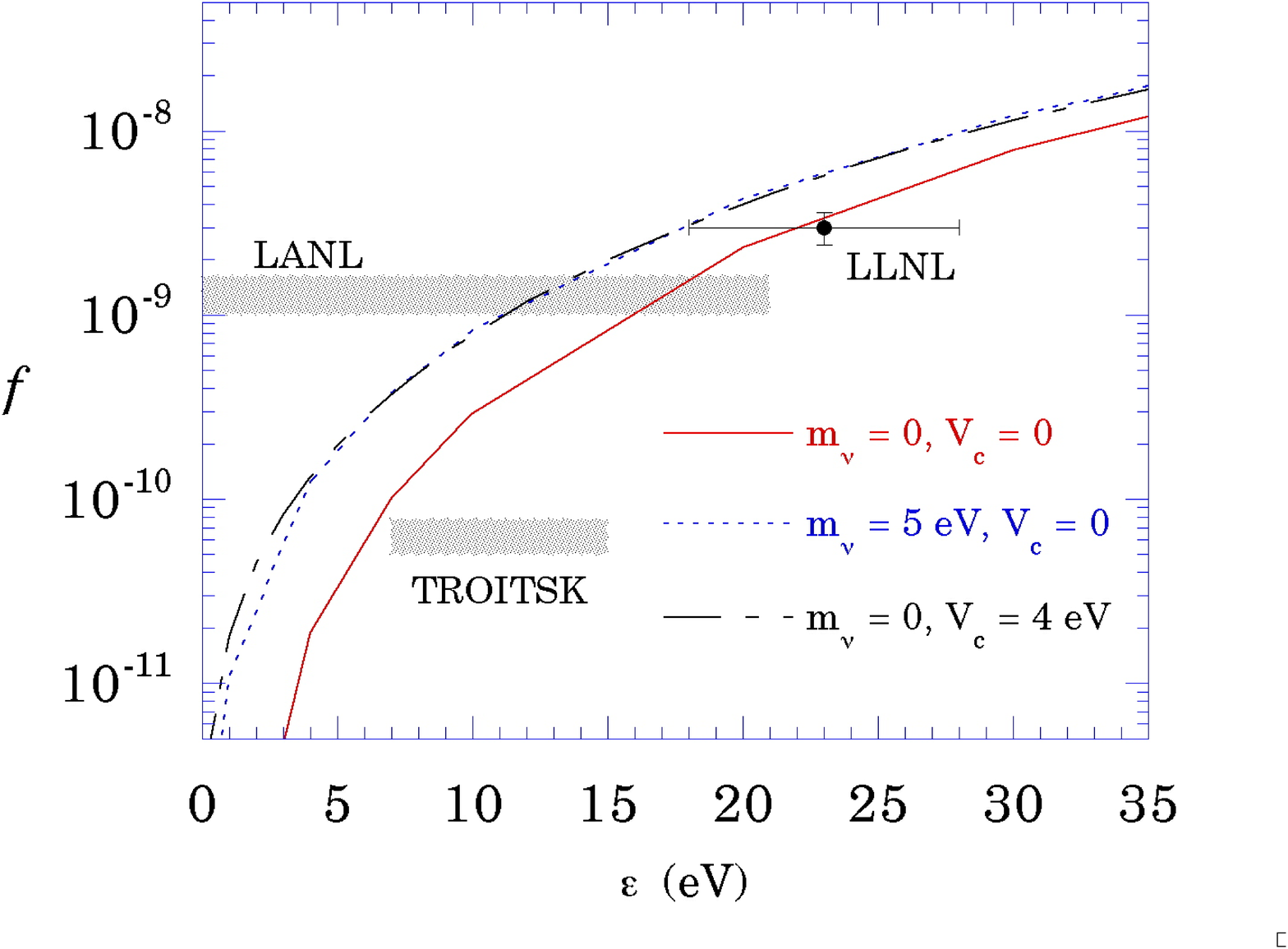}
\caption{Fraction of tritium decays, $f$, with electron kinetic energy
$E_{0}-\varepsilon<E_{e}<E_{0}$, as a function of $\varepsilon$ 
and for different values of the neutrino mass
$m_{\nu}$ and coherent weak potential $V_{c}$ ($E_{0}$= 18.575 keV). The 
line corresponding to  $m_{\nu}=$ 5 eV is shifted to the left by 5 eV so 
that $f(0)=0$. The boxes and dot (see text) correspond to experimental
 best-values for the location and intensity of the spectral 
 excess responsible for the unphysical $m^{2}_{\nu}<0$ values 
 obtained.}
\end{figure}

Such a potential $V_{c}$ has been studied for long, albeit in a 
different context and not yet introduced as a correction 
to $\beta-$decay. (Anti)neutrinos of long-enough deBroglie 
wavelength ($\bar\lambda_{\nu}(cm)=1.97\cdot 10^{-5} / p_{\nu}(eV/c)$), 
when immersed in nuclear matter, cover 
a macroscopic number of nucleons (or quarks) in $\bar\lambda_{\nu}$; hence, 
the collective effect of nuclear matter on the neutrino is 
coherent and averaged over $\bar\lambda_{\nu}$. This is expressed in terms of 
a weak-interaction potential or alternatively as an index of 
refraction associated with the neutrino crossing from one 
material to another. A review of the quantum-mechanical 
principles leading to coherent neutrino scattering is given 
in \cite{14}. This mechanism is behind the proposed methods 
(reviewed in \cite{15}) to detect the relic neutrino background via 
small forces caused by their reflection and refraction 
in target materials. More recently, Loeb \cite{16} has employed 
this potential to show that supernova neutrinos emitted 
with  $E_{\nu}\lesssim$ 50 eV must remain bound to the remnant neutron star. 
Using his notation, $V_{c}(eV)\simeq -3.8\cdot 10^{-14} K \rho_{n}$, 
where $\rho_{n}$ is the density of nuclear 
matter in g/cm$^{3}$,  
$K=\pm \frac{1}{2}\left( 1+\ell~\frac{m_{\nu}}{E}\right)$, 
and $E$ is the total neutrino energy. The upper 
sign in $K$ is for neutrinos with helicity, $-\ell$ and the lower 
sign for antineutrinos with helicity, $\ell$  ($\ell=\pm 1$). 
It must be kept in mind that for nonrelativistic 
Majorana neutrinos, $K\to0$, i.e., 
the presence of the potential can cast 
light on the nature of the 
neutrino emitted. Equivalent expressions for $V_{c}$
can be found in \cite{17}. In $\beta-$decay, the emitted antineutrino 
should therefore experience {\it ab initio} a small repulsive $V_{c}$  
arising from the coherent weak-interaction with the daughter 
atom. This $V_{c}$ is then the traditional potential associated with 
the crossing of a low-energy neutral particle of mass $m$ through 
the boundary between two different materials (daughter nucleus 
and vacuum in this case), therefore changing its momentum, 
$\frac{p^{2}_{2}}{p^{2}_{1}}=1-\frac{2mV_{c}}{p^{2}_{1}}$\cite{17}. 
Taking a representative nuclear radius $R\sim1.2\cdot 10^{-15} 
A^{1/3}$ m ($A$
is the daughter's mass number) yields $\rho_{n}\sim2.3\cdot10^{14}$ 
g/cm$^{3}$ and $V_{c}\sim$ 4.4 eV for 
tritium if $m_{\nu}=$ 0. Since a small variation in the adopted 
$R$ changes $V_{c}$ rapidly, it is sufficient to conclude at this 
point that $V_{c}\sim$ O(1) eV. 

The neutrino total energy and momentum appear explicitly 
in the expression for $P(E_{e})dE_{e}$. The coherent correction is formally 
introduced by making the substitution $E_{\nu} \to E_{\nu}+V_{c}$ in both.
In this regard,
$V_{c}$ is inserted in the same fashion as the Coulomb-screening 
correction \cite{11,13}: $E_{e}$ is shifted by $V_{0}(eV)\simeq\pm 
30.8 ~Z^{4/3}$ (positive sign for positron 
emission, Z being the daughter's atomic number) wherever it 
appears in  $P(E_{e})dE_{e}$ to account for this screening of the Coulomb 
field of the nucleus by the atomic electrons. The classical 
quote by Rose \cite{18} ``the electron distribution is always such 
as though the nucleus were not conscious of the screening 
and as though it emitted electrons into its immediate vicinity 
always in the same way; the only effect of the screening is 
then to accelerate the electrons...'' should apply here with ``$V_{c}$'' 
in place of "the screening" and "antineutrino" as the last word. 
Fig. 2 displays the {\it qualitative}  spectral change due to $V_{c}$ when 
introduced in this fashion, which is precisely an enhancement 
of the expected count rate in the region immediately below $E_{0}$. 
This excess is enticingly similar in magnitude and shape to the 
anomaly in refs. \cite{6,7}. It is also possible to rapidly estimate 
that a coherent potential $V_{c}\sim$ O(1) eV is indeed able to produce a 
{\it quantitative} effect on the electron spectrum
equivalent to that coming from the experimentally-evaluated negative 
$m^{2}_{\nu}$: 
as mentioned, $P(E_{e})dE_{e}$ is proportional to 
$E_{\nu}p_{\nu}=(E_{\nu}+V_{c})\sqrt{(E_{\nu}+V_{c})^{2}-m^{2}_{\nu}}$; 
a numerical value $V_{c}\sim(3/4)^{1/4}\sqrt{\vert m^{2}_{\nu}\vert}$ is 
then seen to drive the magnitude of $P(E_{e})dE_{e}$ similarly in the limits\newline
\newline
i) $V_{c}\to 0$, neutrino kinetic energy $\simeq m_{\nu}$\newline   
and\newline
ii) $m_{\nu}\to 0$, neutrino kinetic energy $\simeq V_{c}$,\newline
\newline
i.e., close to the endpoint in both scenarios. 
Using the weighted average result of 
all tritium experiments $m^{2}_{\nu}=-27\pm20 ~eV^{2}$\cite{19} 
in 
the obtained relation between   
$V_{c}$ and $m^{2}_{\nu}$ 
yields $V_{c}=$ 4.8 eV, in good agreement with the expectation for tritium described 
above; $V_{c}$ as computed should be able to provide a good fit to
the experimental 
data at hand.
Separately and last, it must be mentioned that the fraction 
$f(\varepsilon)$ is not largely changed by the introduction of $V_{c}$ (dash-dot line 
in fig. 1).

The inclusion of this correction in the analysis of present 
tritium experiments may hopefully recover positive values for  
$m^{2}_{\nu}$ 
and improve existent limits; a possible common agreement 
on a $m^{2}_{\nu}>0$ best-value opens up as an exciting possibility. 
The fact that most of these experiments seem to be already 
sensitive to an effect of O(1) eV is encouraging. 	
\begin{figure}[tbp]
\epsfxsize = \hsize \epsfbox{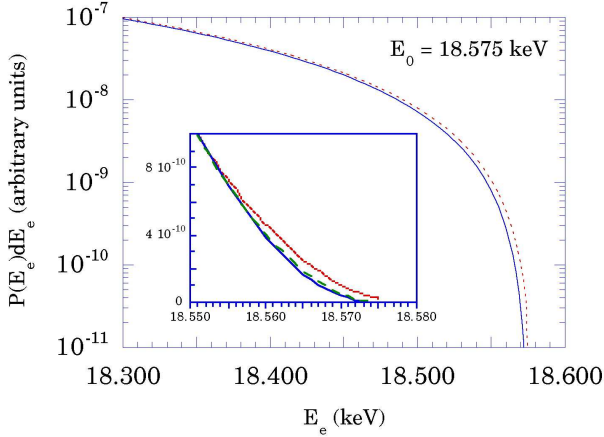}
\caption{Tritium theoretical  $\beta-$spectrum for $m_{\nu}=$ 0 and a coherent 
weak potential $V_{c}=0$  (solid lines), $V_{c}=1~eV$ (dashed) 
and $V_{c}=4~eV$ (dotted). The $V_{c}=0$  and $V_{c}=1~eV$ lines cannot 
be differentiated in the log plot. The insert is a linear blow-up 
of the endpoint region, with all lines normalized to the same value 
at $E_{e}=$ 18.550 keV. The effect of including the coherent 
potential correction is similar in shape and magnitude to 
the endpoint anomaly observed in refs. [5-7,9].}
\end{figure}

{\it Note added July `99:} 
Some time after the first posting of this preprint, H. Terazawa 
kindly called my attention to his early work 
on the neutral current effect in 
$\beta -$decay \cite{terazawa1}. Recently he has revisited
this topic\cite{terazawa2}, arriving at a value $V_{c}=4.71~eV$ 
for tritium and conclusions similar 
to those expressed here. After many instrumental improvements,
the anomalous $m^{2}_{\nu}<0$ remains present in the latest data from the 
Mainz and Troitsk spectrometers \cite{nu98}. To the knowledge 
of this author, these groups have not yet attempted
to interpret their results in the framework of the present discussion, 
favoring instead more contrived explanations\cite{nu98}.
\newline

{\it Acknowledgements:} 
I am indebted to T. Girard for many exchanges on the subject, 
to the Group de Physique des Solides at UniversitŽ Paris VII 
for their hospitality during the 95-96 academic year, and to 
D. Wyler for a critical reading of the manuscript.

\end{document}